\documentclass[11pt]{article}
\usepackage{xspace}
\usepackage{graphicx}
\usepackage{amsmath}
\usepackage{amssymb}
\usepackage{color}

\def\CP{$ C \! P$ }

\def\ra{\rightarrow}

\newcommand\T{\rule{0pt}{2.6ex}}       
\newcommand\B{\rule[-1.2ex]{0pt}{0pt}} 

\definecolor{Red}{rgb}{1,0,0}
\definecolor{Green}{rgb}{0,1,0}
\definecolor{Blue}{rgb}{0,0,1}
\definecolor{Black}{rgb}{0,0,0}

\def\beq{\begin{equation}}
\def\eeq#1{\label{#1}\end{equation}}
\def\eeqn{\end{equation}}

\def\beqa{\begin{eqnarray}}
\def\eeqa#1{\label{#1}\end{eqnarray}}
\def\eeqan{\end{eqnarray}}

\let\bar=\overbar

\def\Dslash{\not{\hbox{\kern-4pt $D$}}}
\def\dslash{\not{\hbox{\kern-2pt $\del$}}}

\def\msb{{\bar{\ssstyle M \kern -1pt S}}}

\def\Title#1{\begin{center} {\Large {\bf #1} } \end{center}}

\begin{document}

\Title{Global Fits of the CKM Matrix with the SCAN Method}

\bigskip\bigskip


\begin{raggedright}  

Gerald Eigen\index{Eigen, G.}, {\it University of Bergen} \\
Gregory Dubois-Felsmann\index{Dubois-Felsmann, G.}, {\it SLAC} \\
David G. Hitlin\index{Hitlin,D.G.} and Frank C. Porter\index{Porter, F.C.}, {\it Caltech}

\begin{center}\emph{}\end{center}
\end{raggedright}

{\small
\begin{flushleft}
\emph{To appear in the proceedings of the 50 years of CP violation conference, 10 -- 11 July, 2014, held at Queen Mary University of London, UK.}
\end{flushleft}
}

\begin{abstract}
We present a Scan Method analysis of the allowed region of the $\bar \rho$ - $\bar \eta$ plane using the latest input measurements of CKM matrix elements,  $\sin 2\beta$, $B^0_{d,s}$ mixing, $\epsilon_K$, $\alpha$ and $\gamma$. In this approach, we make no assumptions as to the distribution of theory uncertainties; rather, we scan over the range of plausible theoretical uncertainties and determine confidence level contours in the $\bar \rho$ -$\bar \eta$ plane. We determine $\alpha$ from branching fraction and \CP\ asymmetry measurements of $B$ decays to all light pseudoscalar-pseudoscalar, pseudoscalar-vector, vector-vector, and $a_1$-pseudoscalar mesons and determine $\gamma$ from $D^{(*)}K^{(*)},~ D^{(*)} \pi$ and $D \rho$ modes, thereby including correlations between the angles of the unitarity triangle. We parameterize the individual decay amplitudes in terms of color-allowed tree, color-suppressed tree, gluing penguin, singlet penguin, electroweak penguin, as well as $W$-exchange and $W$-annihilation amplitudes. Our procedure accounts for all correlations among the fitted CKM parameters $(\bar \rho, \bar \eta, A$ and $\lambda$). The data are consistent with the Standard Model with no need for new physics contributions. We also examine example Òwall plotsÓ, i.e., projections of sensitive parameters showing correlations among them and regions of preferred theoretical parameters.
\end{abstract}

\section{Introduction}

The phase of the CKM matrix is responsible for \CP violation in the Standard Model (SM)~\cite{Kobayashi:1973fv}. 
Unitarity relations of the CKM matrix provide an excellent laboratory to test this prediction. The relation 
$V^*_{ub} V_{ud} + V^*_{cb} V_{cd} + V^*_{tb} V_{td} = 0$ is particularly useful since, in the Wolfenstein 
parametrization~\cite{Wolfenstein:1983yz}, it represents a triangle in the $\bar \rho - \bar \eta$ plane.
For this test, many measurements in the $B$ and $K$ systems can be combined. We perform the test with
the SCAN Method~\cite{Eigen:2013cv}, a frequentist technique in which we make no assumptions as to the distribution of theory uncertainties
of lattice parameters and $|V_{ub}|$ and $|V_{cb}|$; rather, we scan over them 
using grid or MC methods. We determine confidence level (CL) contours in the $\bar \rho - \bar \eta$ plane and
test the hypothesis that the SM is correct using a $\chi^2$ test compared to the alternative that the SM is incorrect.

\section{Fit Methodology}

We perform baseline fits in which we combine measurements of CKM matrix elements, $B \bar B $ mixing, \CP violation in the kaon system and angles of the unitarity triangle 
in the $\chi^2$. 


{\large
\begin{eqnarray}
\label{eq:chisq}
\chi^2(\bar \rho, \bar \eta, p_i, t_j)=
&\Bigl (\frac{\langle  \Delta m_B{_{d,s}}        \rangle -\Delta m_{B_{d,s}}(\bar \rho, \bar \eta, p_i, t_j )} {\sigma_{\Delta m_{B_{d,s}}}}   \Bigr )^2  +
\Bigl ( \frac{\langle | V_{cb,ub,ud,us} |  \rangle - |V_{cb,ub,ud,us}| (\bar \rho, \bar \eta, p_i, t_j ) } {\sigma_{|V_{cb,ub,ud,us}|}}   \Bigr)^2 \nonumber \\
+&\Bigl ( \frac{\langle | \epsilon_K |           \rangle - \epsilon_K  (\bar \rho, \bar \eta, p_i, t_j)} {\sigma_{\epsilon_K}}                 \Bigr)^2 +
\Bigl ( \frac{\langle S_{\psi K^0}            \rangle - \sin2 \beta (\bar \rho, \bar \eta, p_i)} {\sigma_{S_{\psi K^0}}}                        \Bigr)^2 +
\Bigl ( \frac{\langle \alpha                       \rangle - \alpha(\bar \rho, \bar \eta, p_i)} {\sigma_\alpha}                                                   \Bigr)^2  \nonumber \\
+&\Bigl ( \frac{\langle \gamma                    \rangle - \gamma(\bar \rho, \bar \eta, p_i)} {\sigma_\gamma}                                              \Bigr)^2 +
\sum_k \Bigl ( \frac{\langle {\cal M}_k   \rangle - {\cal M}_k (p_i)} {\sigma_{{\cal M}_k}}                                                                    \Bigr)^2 +
\sum_n  \Bigl ( \frac{\langle {\cal T}_n  \rangle - {\cal T}_n(p_i, t_j)} {\sigma_{{\cal T}_n}}                                                              \Bigr)^2.
\end{eqnarray}
}


\noindent
Here, ${\cal M}_k$ terms represent other measurements such as charm and top quark masses and ${\cal T}_n$ terms represent lattice parameters that have theory uncertainties. We also perform ``full fits'' in which we use as inputs 
the measurements that determine $\alpha$ and $\gamma$ rather than the derived values of these angles.
 Confidence regions are determined by comparing
$\chi^2$ values with a critical value instead of looking for a change in $\chi^2$. The algorithm for a $(1-\alpha_c)$\footnote{$\alpha_c$ is a value around $5\%$} confidence region in $d$ dimensions of a $p$ dimensional parameter space with $n$ measurements is as follows: We first determine the  acceptance region, at the $\alpha_c$ significance level, by determining the critical
value $\chi_c^2$ such that $ P(\chi^2\ge \chi_c^2; n-p+d|H_0) \ge \alpha_c$, where $H_0$ is the hypothesis of the SM, and $n-p+d$ is the number of degrees of freedom. 
The confidence region is then given by
all those points in the $d$ dimensional parameter subspace for which $\chi^2\le\chi^2_c$, under $H_0$.


\section{Fit Results in the $\bar \rho -\bar \eta$  Plane}

\begin{figure}[!ht]
\begin{center}
\includegraphics[width=0.9\columnwidth]{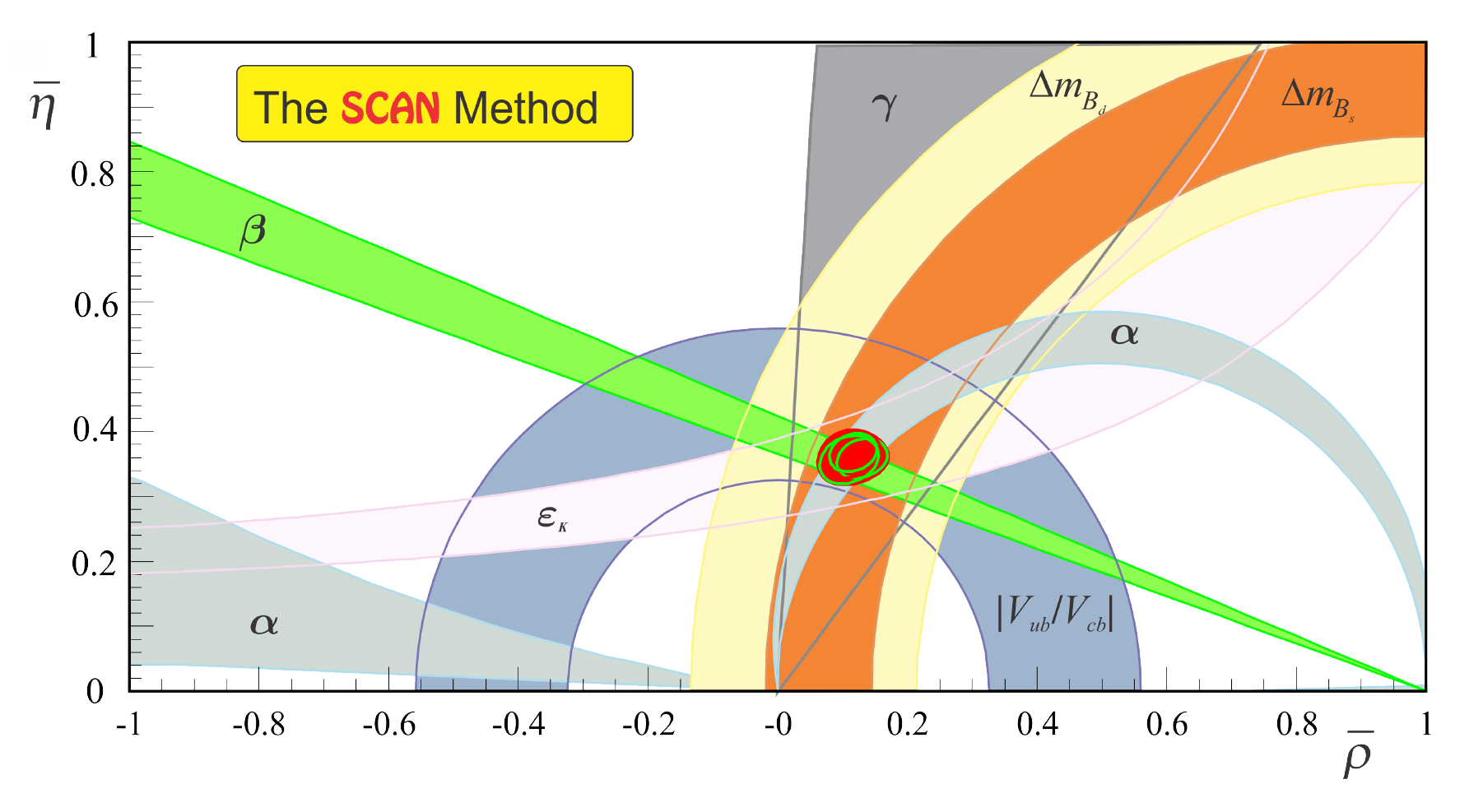}
\includegraphics[width=0.9\columnwidth]{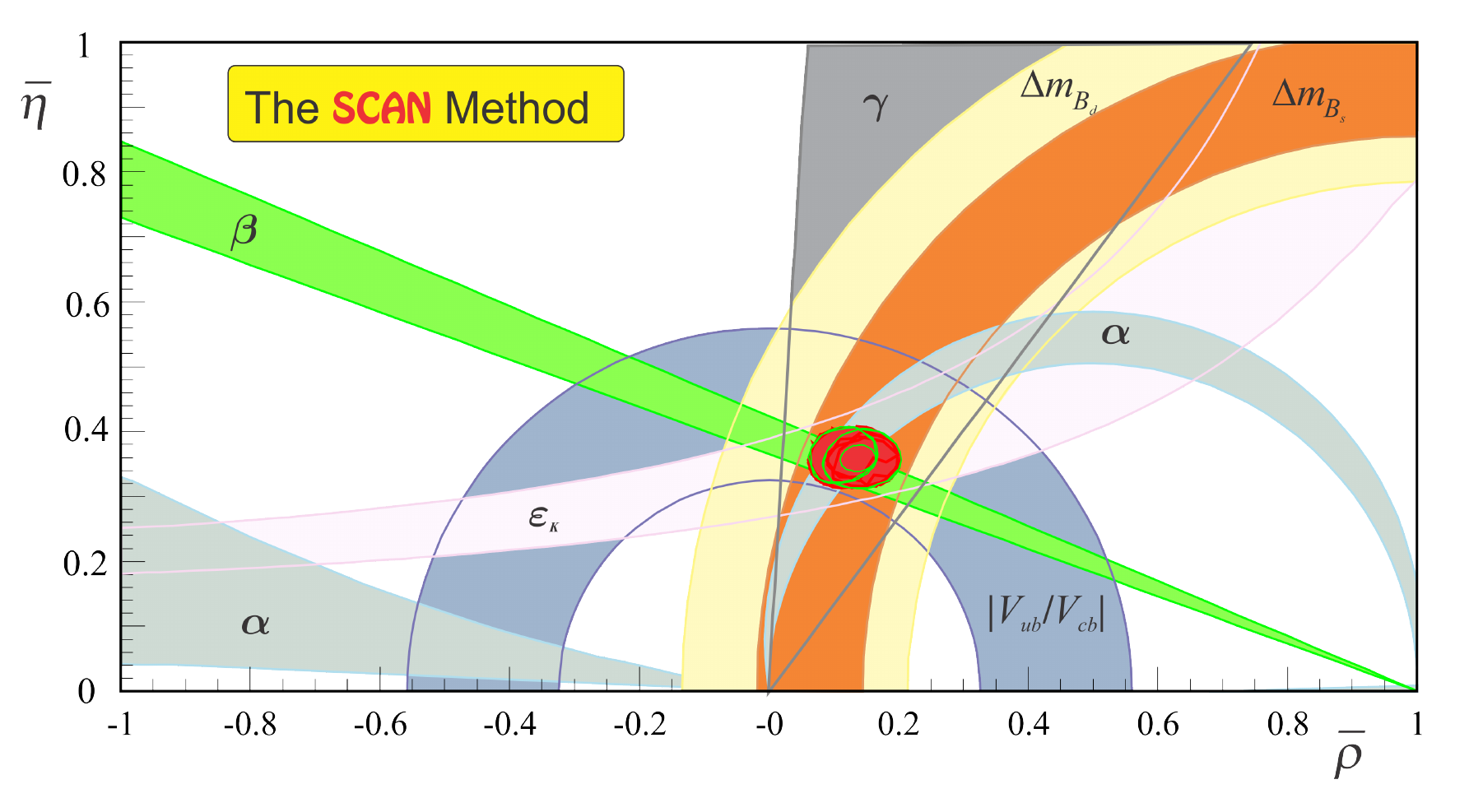}
\caption{Contours at $1-\alpha_c$ CL in the $\bar \rho -\bar \eta$ plane for baseline fits (top) and full fits (bottom).}
\label{fig:baseline}
\end{center}
\end{figure}

In the baseline fits, we fit 23 measurements ($|V_{ud}|, |V_{us}|, |V_{cb}|$,
 $|V_{ub}|, |V_{cs}|, |V_{cd}|, |V_{tb}|$, $\epsilon_K, \Delta m_d$, $\Delta m_s, \sin 2\beta$, $\alpha$\footnote{From a fit to branching fractions and \CP asymmetries of $B \ra PP, PV, VV, a_1P$ decays (see below).}, $ \gamma$\footnote{From a fit  to measurements of $B \ra D^{(*)} K^+, DK^{*+}, D^{(*)} \pi^+, D \rho^+$ decays (see below).}, 
${\cal B}(B\ \ra \tau \nu),  f_{B_s},  f_{B_s}/f_{B_d}, B_{B_s},  B_{B_s}/B_{B_d}, B_K, 
m_t , m_c ,  \tau_{B_d}, \tau_{B_s})$ listed in Tables~\ref{tab:measurement} and
\ref{tab:lattice} to 13  parameters $(\bar \rho, \bar \eta$, $A, \lambda,  
f_{B_s},  f_{B_s}/f_{B_d}, B_{B_s},  B_{B_s}/B_{B_d}, B_K, m_t , m_c ,  \tau_{B_d}, \tau_{B_s})$.
The PDG~\cite{Beringer:1900zz} uses scaling factors of 2.6 and 2.0 for averaging $|V_{ub}|$ and $|V_{cb}|$ results from inclusive and exclusive 
modes, respectively.  This procedure increases all errors.
 Figure~\ref{fig:baseline} (left) shows  
the overlay of $1-\alpha_c$ confidence level (CL) contours of all accepted fits.

\begin{table}[!th]
\begin{center}
\caption{Measurement inputs. }
\begin{tabular}{|l|c|c||l|l|c|c|}  \hline\hline
Input  & Value  & Ref& Input  &  Value  &  Ref \\ \hline
$|V_{cb}|$ & $(4.09\pm 0.069 \pm 0.09)\times10^{-2} $&\cite{Beringer:1900zz} &$m_t$ & $(173.07 \pm 0.88)  \rm GeV/c^2 $&  \cite{Beringer:1900zz} \\
$|V_{ub}|$ & $(4.15 \pm 0.31 \pm 0.39)\times10^{-3} $&\cite{Beringer:1900zz}& $m_c$ & $(1.275 \pm 0.025) \rm GeV/c^2 $ &  \cite{Beringer:1900zz} \\
$|V_{us}|$ &$ 0.2252 \pm 0.0009$ & \cite{Colangelo:2010et}& $ \Delta m_d$ & $ (0.510 \pm 0.003) \rm ps^{-1}$  &  \cite{Beringer:1900zz} \\
$|V_{ud}|$ & $0.97425 \pm 0.00022$ & \cite{Colangelo:2010et}& $ \Delta m_s$ & $ (17.761 \pm 0.022) \rm ps^{-1}$  &  \cite{Beringer:1900zz} \\
$|V_{cd}|$ & $0.23 \pm 0.11$ & \cite{Beringer:1900zz} & $\epsilon_K$ & $ (2.228 \pm 0.0011)\times10^{-3} $ &  \cite{Beringer:1900zz} \\
$|V_{cd}|$ & $1.006±  \pm 0.023$ &  \cite{Beringer:1900zz}& $ \sin 2 \beta$ & $ 0.682 \pm 0.019 $ &  \cite{Beringer:1900zz}  \\
$|V_{tb}|$ & $0.97 \pm 0.08$ &  \cite{Beringer:1900zz}& $\alpha$ & $85.1^{+2.2^\circ}_{-2.0}$ & $^\dagger$ \T \B   \\
${\cal B}(B \ra \tau \nu)$ &$(1.15\pm 0.23) \times 10^{-4}$ &  \cite{Amhis:2012bh} & $\gamma$ & $ 78.9^{+ 6.6}_{-10. 3}$  & $^\ddagger$ \T \B \\ 
 \hline\hline
\end{tabular}
\label{tab:measurement}
\end{center}
\end{table}
\begin{table}[!th]
\begin{center}
\caption{Lattice parameters and QCD parameters.}
\vskip 0.2 cm
\begin{tabular}{|l|c|c||l|c|c|}  \hline\hline
 Lattice & Value & Ref& QCD & Value & Ref  \\ 
 parameter &  & & parameter & & \\ \hline
 $ f_{B_s} $ &$ (228.66\pm 2.0 \pm 5.5) \rm MeV$ &\cite{Laiho:2009eu} & $\eta_{cc}$ & $ 1.39 \pm 0.35 $ & \cite{Gilman:1982ap, Herrlich:1993yv}\\
 $ f_{B_s}/f_{B_d} $ &  $1.205 \pm 0.0086 \pm 0.0188 $ &\cite{Laiho:2009eu}&  $\eta_{tc}$ & $ 0.47±  \pm 0.04 $& \cite{Gilman:1982ap, Herrlich:1996vf} \\
 $ B_{B_s} $ & $ 1.311 \pm 0.046 \pm 0.076$  & \cite{Laiho:2009eu}& $\eta_{tt}$ & $0.5765±  \pm 0.0065 $&\cite{Gilman:1982ap, Buras:1990fn} \\ 
 $ B_{B_s}/B_{B_d} $ & $ 1.053 \pm 0.040 \pm 0.064$  &\cite{Laiho:2009eu}&  $\eta_{b}$  & $0.551 \pm 0.007 $& \cite{Buras:1990fn} \\
 $B_K$ & $ 0.7584  \pm 0.0020 \pm 0.019 $&\cite{Laiho:2009eu} &  & &\\
 \hline\hline
\end{tabular}
\label{tab:lattice}
\end{center}
\end{table}

We also perform ``full fits'' in which we use 256 branching fraction and \CP asymmetry
measurements instead of the $\alpha$ and $\gamma$ direct inputs. These fits, with 114 parameters, uniquely among the procedures in common 
use~\cite{Bona:2007vi, Charles:2004jd} account for possible correlations among $\alpha, \beta$ and $\gamma$.  
We currently include all branching fraction and \CP asymmetry
measurements of $B$ to pseudoscalar-pseudoscalar ($PP$), pseudoscalar-vector ($PV$), vector-vector ($VV$) and $a_1$-pseudoscalar ($a_1 P$)
modes to determine $\alpha$. We include tree, color-suppressed tree, penguin, singlet penguin, $W$-exchange, $W$-annihilation,
and EW penguin amplitudes (up to $\lambda^3$ beyond leading order), as well as $SU(3)$ corrections in defining the amplitudes. We parametrize the
observables using the Gronau-Rosner approach~\cite{Gronau:1994rj, Gronau:1998fn, Dighe:1997wj, Gronau:1999hq, Gronau:1999qd, Beneke:2006rb}.
The amplitude ratios and \CP asymmetry measurements of
 $B^\pm\ra DK^\pm,~ B^\pm \ra D^*K^\pm$ and $B\ra DK^{*\pm}$ modes determine $\gamma$, 
 while those of
$B^\pm \ra  D \pi^\pm, ~B^\pm \ra D^*\pi^\pm$ and $B \ra D\rho^\pm$ modes
 determine $\sin(2\beta +\gamma)$. 
Figure~\ref{fig:baseline} (right) shows an overlay of $1-\alpha_c$~ CL contours of all successful full fits in $\bar \rho -\bar \eta$ plane. 

\vskip -0.5 cm
\begin{table}[!th]
\begin{center}
\caption{$1-\alpha_c$ ~CL ranges for  Unitarity Triangle parameters from baseline fits and full fits.}
\vskip 0.2cm
\begin{tabular}{|l|c|c||c|}  \hline\hline
 Parameter& Baseline Fit & Baseline Fit  & Full Fit  \\ 
  & without $B \ra \tau \nu$ & with  $B \ra \tau \nu$ &  without $B \ra \tau \nu$ \\  \hline
 $ \bar \rho $ &$ 0.069-0.144 $ & $0.073-0.145$ & $ 0.081-0.126$ \\
 $ \bar \eta $ & $0.320-0.395 $ & $0.324-0.396$ & $0.331-0.375$ \\ 
 $ \beta ~[^\circ] $ & $ 19.4-24.2$  &  $19.7-24.3 $ & $20.1-22.8 $\\ 
 $  \alpha ~[^\circ] $ & $ 79.8-90.2 $  &  $80.0-90.1 $  & $81.8-88.2$ \\
 $ \gamma ~[^\circ]$ & $68.0-78.7 $ & $68.0-78.2 $ & $70.3-77.0 $\\
 \hline\hline
\end{tabular}
\label{tab:example}
\end{center}
\end{table}
\vskip -2cm



\section{Extraction of $\alpha$ and $\gamma$}

\begin{figure}[!ht]
\begin{center}
\includegraphics[width=0.48\columnwidth]{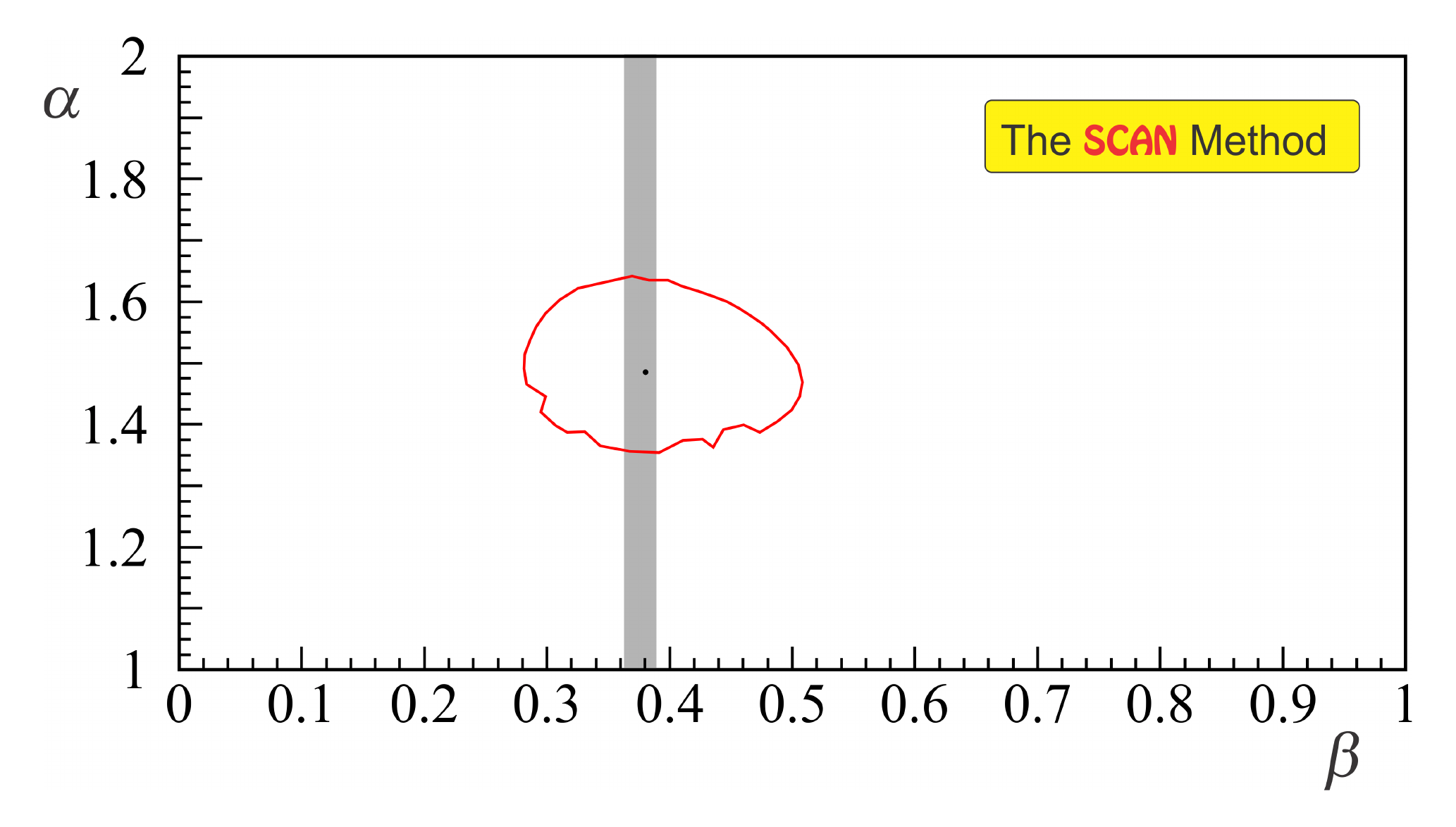}
\includegraphics[width=0.5\columnwidth]{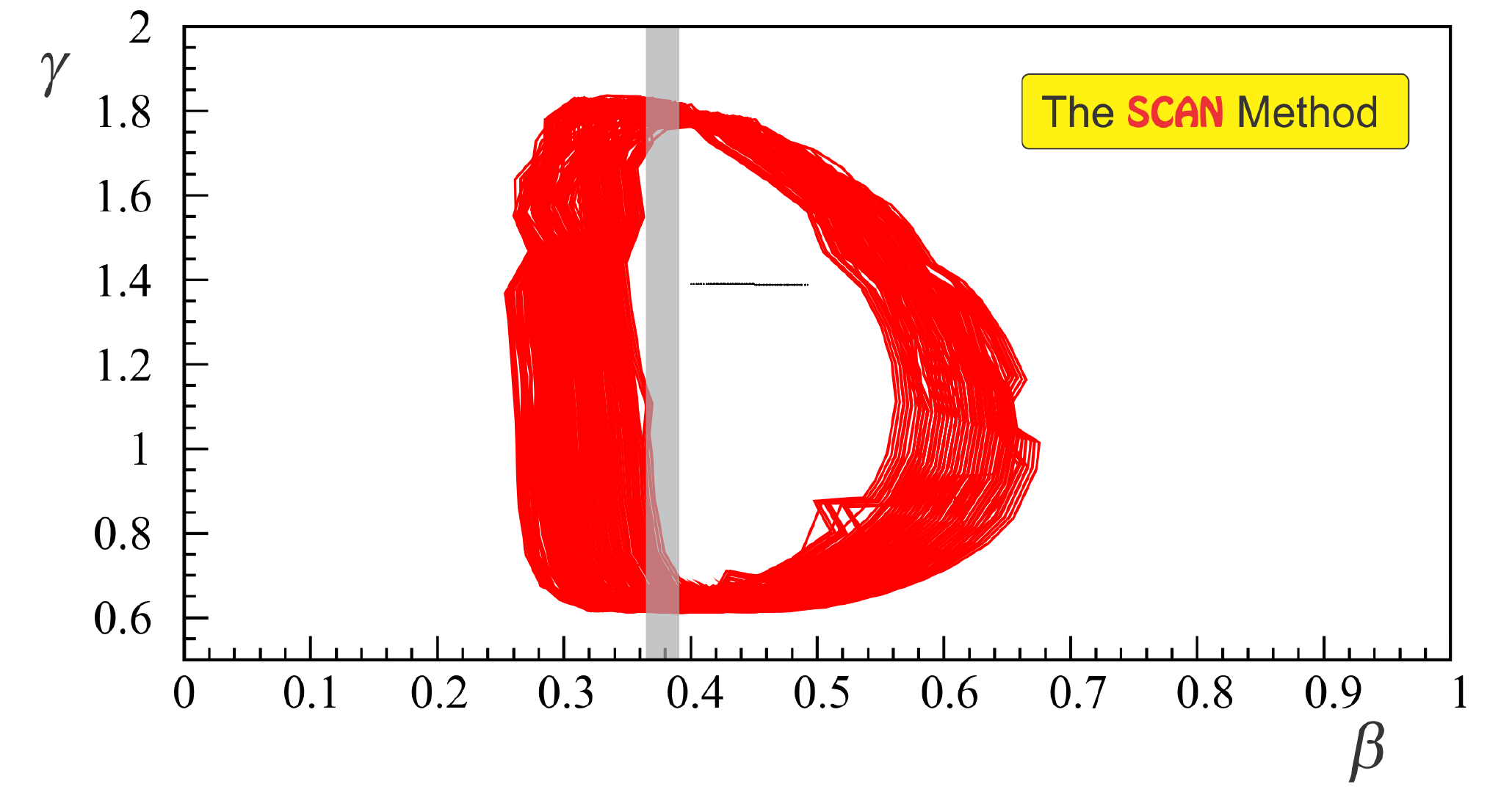}
\vskip -0.2cm
\caption{The $1-\alpha_c$~ CL contour in the $\alpha -\beta$ plane from fits to branching fractions and \CP asymmetries of $B \ra PP, ~PV, ~VV, ~a_1P$ modes (left) and  $1-\alpha_c$~ CL contours in the $\gamma -\beta$ plane from fits to $B^\pm \ra D^{(*)} K^\pm, DK^{*\pm},  D^{(*)} \pi$, and $D^- \rho^+$ modes (right). The individual $\gamma - \beta$ contours result from a scan over
 $|V_{ub}/V_{cb}| $. The grey-shaded band shows the world average of $\beta$ measured in $b \ra c \bar c s$ modes.}
\label{fig:angle}
\end{center}
\end{figure}

We use 181 branching fraction and \CP asymmetry measurements of  $B \ra PP$, $B \ra PV$, $B \ra VV$, and $B\ra a_1P$ modes~ \cite{Amhis:2012bh}to fit 94 parameters and determine the $\alpha-\beta$ contour $@(1-\alpha_c)$  CL  shown in Fig.~\ref{fig:angle} (left). The central value of $\beta$ is consistent with the measured world average of $ \beta  =(21.5^{+0.8}_{-0.7})^\circ$ extracted from $\sin 2 \beta$ measured in $b \ra c \bar c s$ modes. Further, we use 56 branching fraction and \CP asymmetry  measurements of $B^\pm \ra D^{(*)} K^\pm,  DK^{*\pm}, D^{(*)} \pi$, and $ D^- \rho^+$ modes~ \cite{Amhis:2012bh} analyzed with the GLW~\cite{Gronau:1990ra, Gronau:1991dp}, ADS~\cite{Atwood:1996ci, Atwood:2000ck} and GGSZ~\cite{Giri:2003ty} methods to fit  19 parameters and plot $\gamma -\beta$ contours \ $@(1-\alpha_c)$~ CL, which are shown in Fig.~\ref{fig:angle} (right).  Since the contours depend on $|V_{ub}/V_{cb}|$, we explicitly scan over the ratio.  The central values for $\beta$ again agree with the world average measured in  $b \ra c \bar cs$ decay modes.

\begin{table}[!th]
\begin{center}
\caption{Determination of the angle $\alpha$ from fits to branching fractions and \CP asymmetries of $B \ra PP,~ PV, ~ VV$,  and $a_1 P$ modes and determination of $\gamma$ from rate and \CP asymmetry measurements of $B^\pm \ra D^{(*)} K^\pm$,
 $B^\pm  \ra DK^{*\pm}$, $B \ra D^{(*)} \pi$, and $B^0 \ra D^- \rho^+$ modes.}
\vskip 0.3 cm
\begin{tabular}{|l|c|c|c|}  \hline\hline
 Mode & $\alpha [^\circ]$ & $ \beta[^\circ]$  &$\gamma[^\circ]$ \\ \hline
 $ B \ra PP+PV+ VV+ a_1 P$ & $ 85.1^{+2.2}_{-2.1}  $  &  $21.8^{+1.6}_{-2.1}$  & \T \B \\ \hline
 $ B \ra D^{(*)} h$ &  & $22.8^{+7.7}_{-2.0} $ & $78.9^{+6.6}_{-10.3}$ \T \B \\
 \hline\hline
\end{tabular}
\label{tab:example}
\end{center}
\end{table}

\section{Wall Plots}

\begin{figure}[!ht]
\begin{center}
\includegraphics[width=0.48\columnwidth]{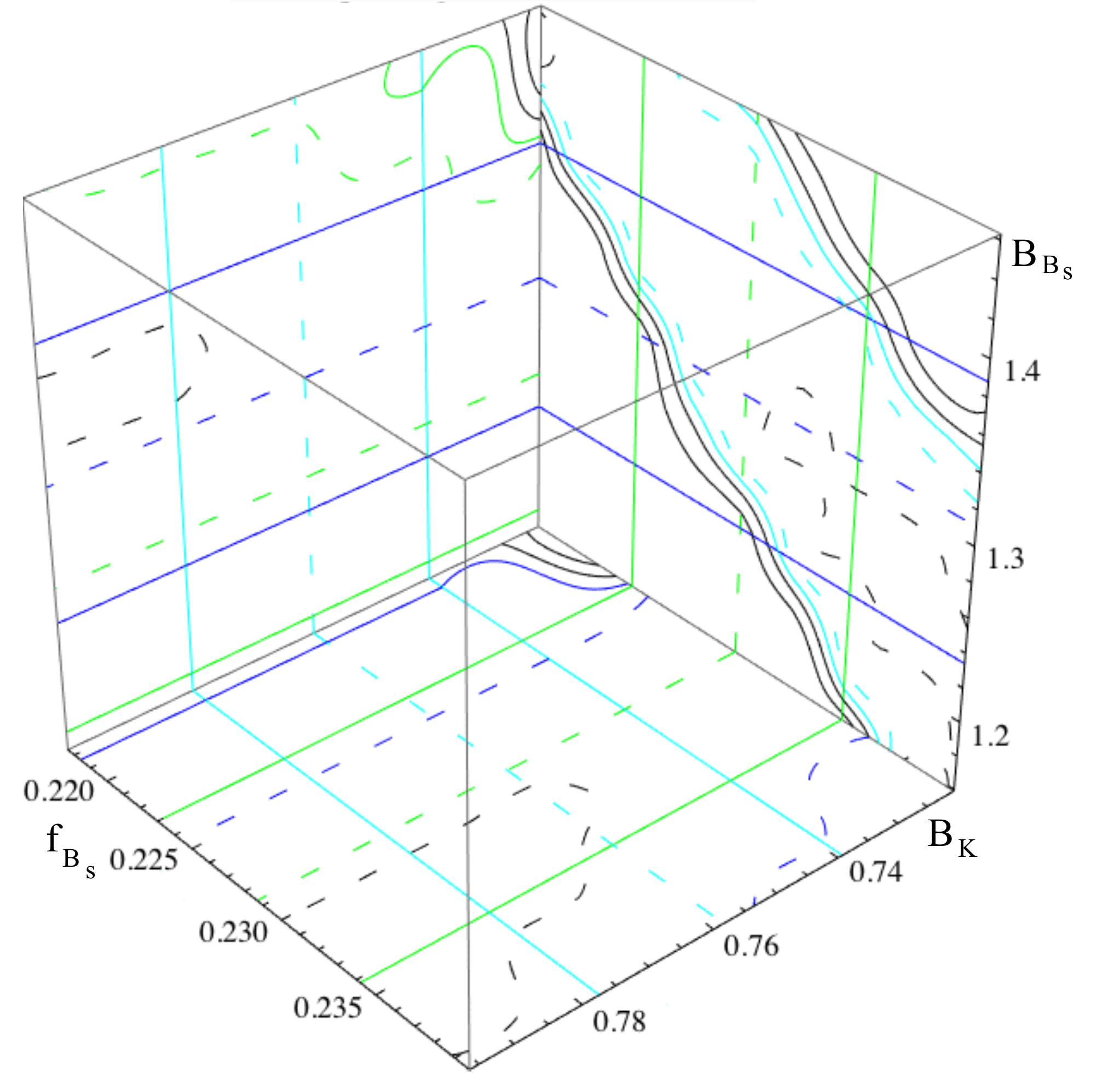}
\includegraphics[width=0.48\columnwidth]{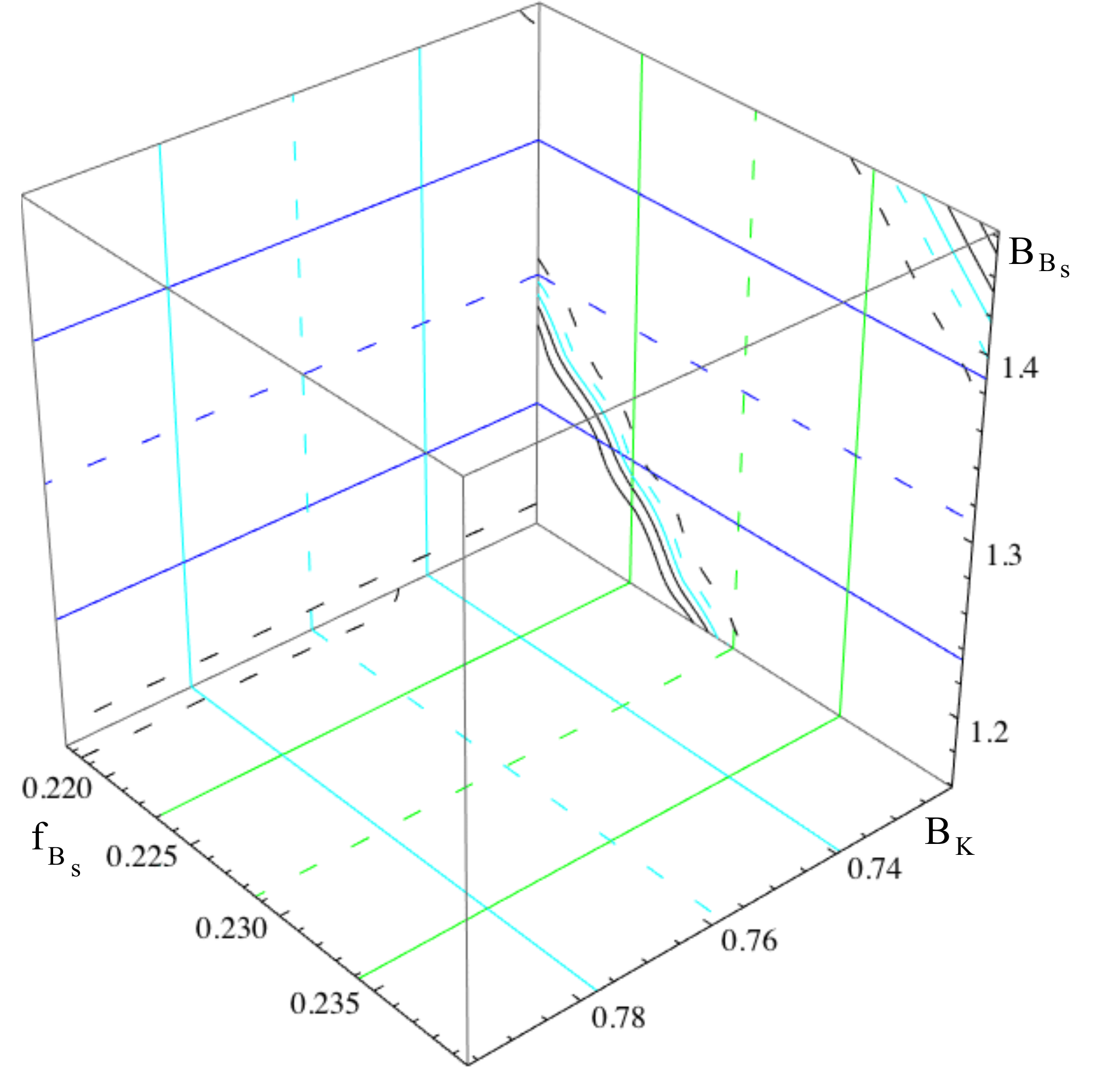}
\includegraphics[width=0.48\columnwidth]{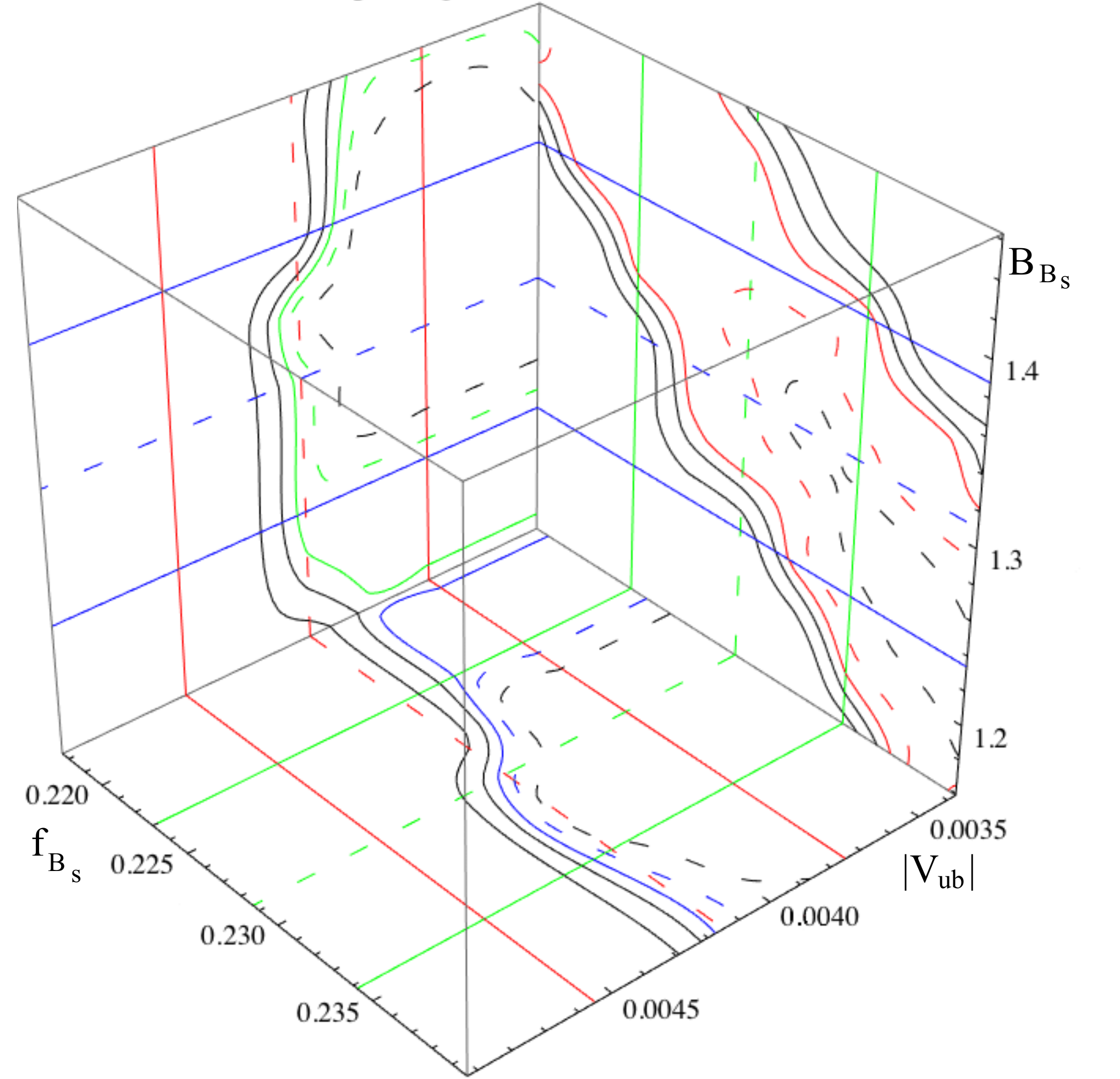}
\includegraphics[width=0.48\columnwidth]{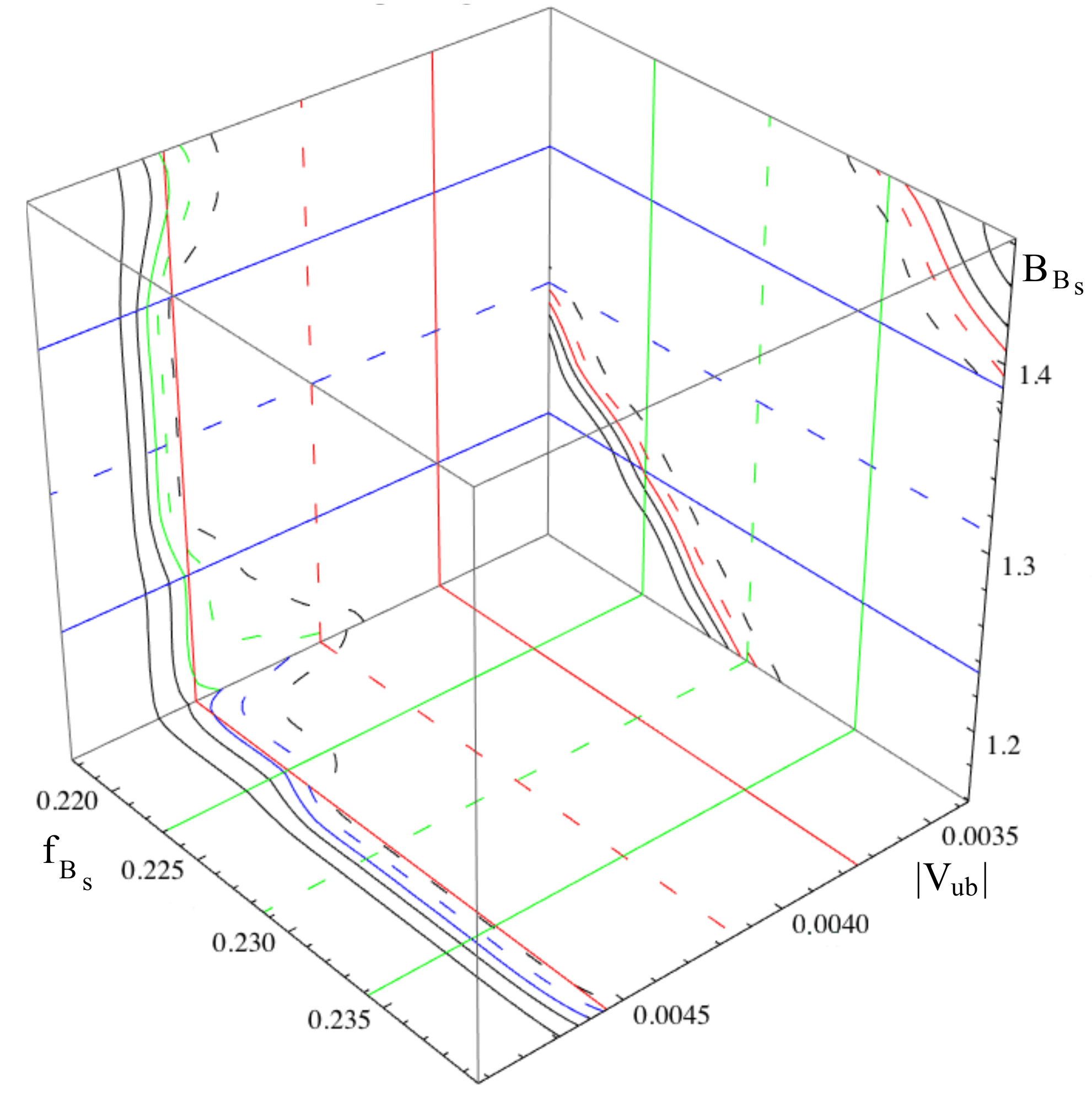}

\caption{Wall plots for $B_{B_s}$ versus $B_K$ versus $f_{B_s}$ (top) and $B_{B_s}$ versus $|V_{ub}|$ versus $f_{B_s}$ (bottom) for $68\%$ contours (left) and $1-\alpha_c$ contours (right).}
\label{fig:wallplot}
\end{center}
\end{figure}

We construct wall plots to display the  correlations among sets of three out of the seven parameters with large theory uncertainties. To study the impact of the remaining parameters on two parameters displayed,  we impose different constraints on the other displayed and undisplayed parameters. These studies clearly show that certain regions of the theory parameters are favored by the data while others are not.  As an example,
Figure~\ref{fig:wallplot} shows correlations for $B_{B_s}$ versus $B_K$ versus $f_{B_s}$ (top) and  $B_{B_s}$ versus $|V_{ub}|$ versus $f_{B_s}$ (bottom) for $68\%$ CL (left) and $1-\alpha_c$ CL (right). Orthogonal solid lines in each plane show the $\pm 1\delta$  theory error range. Outer black contours result from a probability requirement of $>\!32\%$ or $>\!5\%$. Inner black contours result from a  $\pm 1\delta$ requirement on all undisplayed parameters while colored solid contours result from a  $\pm 1\delta$ requirement on the out-of-plane variable.  Constraining the out-of-plane variable to its central value yields the colored dashed contours while the black dashed contours result from constraining undisplayed variables to their  central values. The plots show that larger values of $B_K$ and $f_{B_s}$ and smaller values of $|V_{ub}|$ are favored.

\section{Conclusion}

Baseline fits and full fits performed with the SCAN method yield results that are in good agreement with the SM without making any assumptions about the distribution of theory uncertainties as is done for $\rm UT_{fit}$~\cite{Bona:2007vi} and CKMfitter~\cite{Charles:2004jd} analyses. The full fits how a slightly larger allowed region in the $\bar \rho -\bar \eta$ plane. They account for correlations among the observables $\alpha$,  $\beta$ and $\gamma$.
These correlations, however, are ignored in $\rm UT_{fit}$ and CKMfitter analyses. 
Wall plots allow the determination of whether any potential SM discrepancy originates from  the  values of theory parameters or from experimental measurements. The measurements prefer larger values of $B_K$ and $f_{B_s}$ and smaller values of $|V_{ub}|$.




\bigskip
\section{Acknowledgments}

This work is supported in part by the U.\ S.\ Department of Energy under Grant DE-FG02-92-ER40701
and by the NFR (Norway).

%
%

%
%
%
%
 
\end{document}